\documentclass[a4paper,fleqn]{cas-sc}

\usepackage[authoryear]{natbib}
\usepackage{booktabs} 

\usepackage{lineno}

\usepackage[utf8]{inputenc}
\DeclareUnicodeCharacter{2206}{\unicode{2206}}

\def\tsc#1{\csdef{#1}{\textsc{\lowercase{#1}}\xspace}}
\tsc{WGM}
\tsc{QE}
\tsc{EP}
\tsc{PMS}
\tsc{BEC}
\tsc{DE}

\begin{document}
\let\WriteBookmarks\relax
\def\floatpagepagefraction{1}
\def\textpagefraction{.001}



\title [mode = title]{Electric Vehicle Sales Forecasting Model Considering Green Premium: A Chinese Market-based Perspective}                      

\tnotetext[1]{This document is the results of the research
   project funded by the National Science Foundation.}

\tnotetext[2]{The second title footnote which is a longer text matter
   to fill through the whole text width and overflows into
   another line in the footnotes area of the first page.}

%
\author[1]{Zhi Li}

\fnmark[1]


\address[1]{organization={PBC School of Finance},
    addressline={Tsinghua University}, 
    city={Beijing},
    postcode={100084}, 
    country={China}}

\author[1]{Hang Fan}[type=editor,
                        auid=000,bioid=1]
\cormark[1]
\ead{fanhang123456@163.com}

\author[2]{Shuyan Dong}[type=editor,
                        auid=000,bioid=1]
\address[2]{organization={Institute for Interdisciplinary Information Sciences},
    addressline={Tsinghua University}, 
    city={Beijing},
    postcode={100084}, 
    country={China}}
\ead{dsy17@mails.tsinghua.edu.cn}




\fntext[fn1]{This is the first author footnote. but is common to the third author as well.}
\fntext[fn2]{Another author footnote, this is a very long footnote and
  it should be a long footnote. But this footnote is not yet
  sufficiently long enough to make two lines of footnote text.}

\nonumnote{This note has no numbers. In this work, we demonstrate $a_b$
  the formation Y\_1 of a new type of polariton on the interface
  between a cuprous oxide slab and a polystyrene micro-sphere placed
  on the slab.
  }

\begin{abstract}
"Green Premiums" which means the difference in cost between emissions-emitting technology and zero-emissions or emissions-reducing technology is significant for those renewable energy technology to address the climate change challenge facing the world in this century. China's Electrical Vehicles (EVs) industry is the first to cross the green premium into the commercialization stage, prompting its market size to exceed that of the US and EU combined, making it the most inspiring case in global carbon reduction practices. This study, which is based on first-hand data from industry research and innovatively constructs a multi-factor green premium model for EVs based on Total Cost of Ownership (TCO) analysis, finds that EVs currently have a higher green premium than Internal Combustion Engine Vehicles (ICEVs) and is expected that short-range EVs will be the first to achieve parity in acquisition costs by 2025 and long-range EVs by 2030. Further, this paper constructs a generalized Bass diffusion model considering the green premium, selects the time series data of EVs diffusion in the Chinese market from 2010-2021, and uses a genetic algorithm to fit the parameters to predict the EVs market penetration in the next ten years under different scenarios. The model prediction results show that EVs are successful innovative diffusion products and the market penetration rate depends largely on their green premium. The Chinese EVs market may experience a slowdown or even a decline in growth in the short term, but will maintain high growth in the medium to long term, with annual sales expected to reach 10.77 million units by 2030 and a penetration rate of about 39\%. 
\end{abstract}


\begin{highlights}
\item The economic modeling of the green premium of EVs by disaggregating the cost components and the full life cycle cost of EVs; 
\item The impact of the green premium of EVs on the Bass model based on the generalized Bass model;
\item The use of the genetic algorithm to fit the Bass model with the green premium to forecast the Chinese EV market based on future scenarios. 
\end{highlights}

\begin{keywords}
Green Premium\sep Electric Vehicle \sep Market forecasting \sep Bass Model \sep Chinese Market
\end{keywords}

\maketitle

\section{Introduction}

Green Premiums, the difference in cost between emissions-emitting and zero-emissions or emissions-reducing technology, was introduced by Bill Gates in his book The Climate Economy and the Future of Humanity as a central grip on the economy and industry to achieve carbon neutrality \citep{gates2021avoid}. The green premium extrapolates from the perspective of technological innovation from near-to-far to the gradual substitution of emissions-emitting technology for zero-emissions technology. This concept is further substantiated considering the International Energy Agency (IEA) also uses the concept of energy transition costs (i.e., green premium) to describe the path to carbon neutrality.\par

The persistence of the green premium is due to the inability of economic sectors to overcome market failures in the process of low-carbon transition, including factors such as negative environmental externalities, RD spillovers, and monopolistic market power. While fossil energy has developed over centuries to form both a mature global energy supply system and internal combusition engine vehicle production line, the new energy vehicle market is also characterized by high technology and supply chain barriers, large initial capital requirements, and long payback periods, thereby resulting in the automotive industry stagnating in its development of innovative energy technologies over the past century. \par

The transportation sector has long been one of the major emitters of GHGs (accounting for about 20\% of total anthropogenic GHG emissions), as well as the emergence of New Energy Vehicles (NEVs), mainly Electric Vehicles (EVs), has effectively reduced the terminal carbon emissions of Internal Combustion Engine Vehicles (ICEVs), effectively overcoming market failures and significantly reducing the green premium of the industry, setting the first major precedent of any  traditional industry to reduce emissions. As the world’s largest automotive market, China’s EV market has entered an explosive growth phase in the last decade and its electrification may have a significant impact on the global transportation energy transition. This transition inevitably brings out most illuminating empirical data in mitigating the energy crisis. However, past academic studies have not explained the process of how technological change in EVs affects industry penetration in economic terms, whereby there is a disconnect between research on the green premium and market capacity of EVs. \par

First, studies on the factors influencing the green premium in the electric vehicle industry can be divided into two main categories: supply-side and demand-side. At the supply-side regulatory level, the factors that influence the green premium of electric vehicles include fuel economy standards \citep{al2014analysis}, credit trading system \citep{wang2018impacts}, and greenhouse gas emission regulations \citep{panwar2018analysis}. The influencing factors on the demand side can be divided into two main categories. The first category is mainly the economic cost of EVs, including EV prices and government subsidies \citep{adepetu2017relative}, the shadow value of car licenses \citep{kong2020effects}, and fuel and alternative fuel costs \citep{javid2017comprehensive}. At the usage level, the driving experience of EVs \citep{skippon2016experience}, range \citep{lee2017efficiency}, energy efficiency \citep{rogers2003empowerment}, charging time, and charging facility density \citep{zou2020effects} can have a significant impact on the green premium of EVs. However, most studies on the green premium of EVs are limited to the acquisition cost of EVs and only correlate the selling price with sales volume, without considering the cost of the full life cycle of the vehicle as a durable good and considering the impact of comprehensive factors on the market such as the influence of government policies.\par

Secondly, to accurately forecast the EV market volume, it is crucial to choose an appropriate model to forecast EV market sales. The current models for market demand forecasting mainly include the consumer intention survey method \citep{juster1966consumer}, integrated salesperson opinion method \citep{peterson1989sales}, expert opinion method, market test method, time series analysis method \citep{ansuj1996sales}, linear trend method, statistical demand analysis method \citep{bicheng2018}, and so on. The Bass diffusion model is also considered to be a typical method for predicting the market penetration of EVs \citep{renbin@2013diffusionmodel}, and since this paper precisely uses the EV industry as a representative sample to study the sales of an industry where a green premium exists, the Bass diffusion model is well suited for the current scenario.\par

In view of this, this paper combines vehicle engineering and economics to construct a bottom-up economic model of the green premium for EVs by disaggregating the cost components and the life-cycle costs of EVs using a combination of qualitative and quantitative methods. This economic model, combined with a modified Bass diffusion model, is used to conduct a future scenario forecast analysis of the Chinese EV market. The innovations of this paper include:\par
(1) We built up an economic model of the green premium between emissions-emitting technology and zero-emissions technology, offering a technical perspective to understand the full picture and the current stage of Carbon Neutral. The green premium of EVs is defined by disaggregating the cost components and the full life cycle cost, and the primary data on the automotive industry for the model is collected by field investigations and expert interviews. Based on that, green premium can be defined in both the quantitative and qualitative aspects, allowing for a more accurate way to observe and predict green technology’s commodification.

(2) We modified the generalized Bass model with the variable “\textbf{green premium}” to predict the sales market of the EVs, which improve the accuracy of Bass model, enhancing the $R^{2}$ from 0.6 to 0.77. The Bass modeling method is highly adaptable and is easy to be expanded to other green innovative product market sales prediction scenario such as photovoltaic, wind power and so on.

(3) We creatively used the genetic algorithm to fit the modified Bass model so that the most suitable model parameter can be selected. In the case study section, the modified prediction method is tested on the Chinese EV market according to the annual data from 2010 to 2021 and the prediction results proved its effectiveness. 

The paper is organized as follows: Section 2 models the economics of the green premium, Section 3 develops a generalized Bass diffusion model considering the green premium, and Section 4 calculates the green premium for EVs in the Chinese market based on the actual data collected from the research and forecasts of the EV sales. Section 5 presents the related conclusions. 

\section{Economic modeling of the green premium based on total cost of ownership analysis}
TCO refers to the total costs incurred during the entire life cycle of a product, from initial design and manufacturing to shipping, usage, and disposal\citep{ellram1993total}. Engineers commonly use TCO analysis to determine the total cost of producing a system or product, and it is a key indicator of product competitiveness. This paper utilizes the TCO framework to develop a green premium model for EVs.

In this paper, the TCO cost model for automobiles is built based on automotive engineering. The green premium for EVs is divided into four parts: production cost (supply side), purchasing cost (demand side), operating cost associated with its ownership, and the residual value of the used vehicles. The specific cost components are shown in the figure below.
\begin{figure}[ht]
    \centering
    \includegraphics[width=0.8\textwidth]{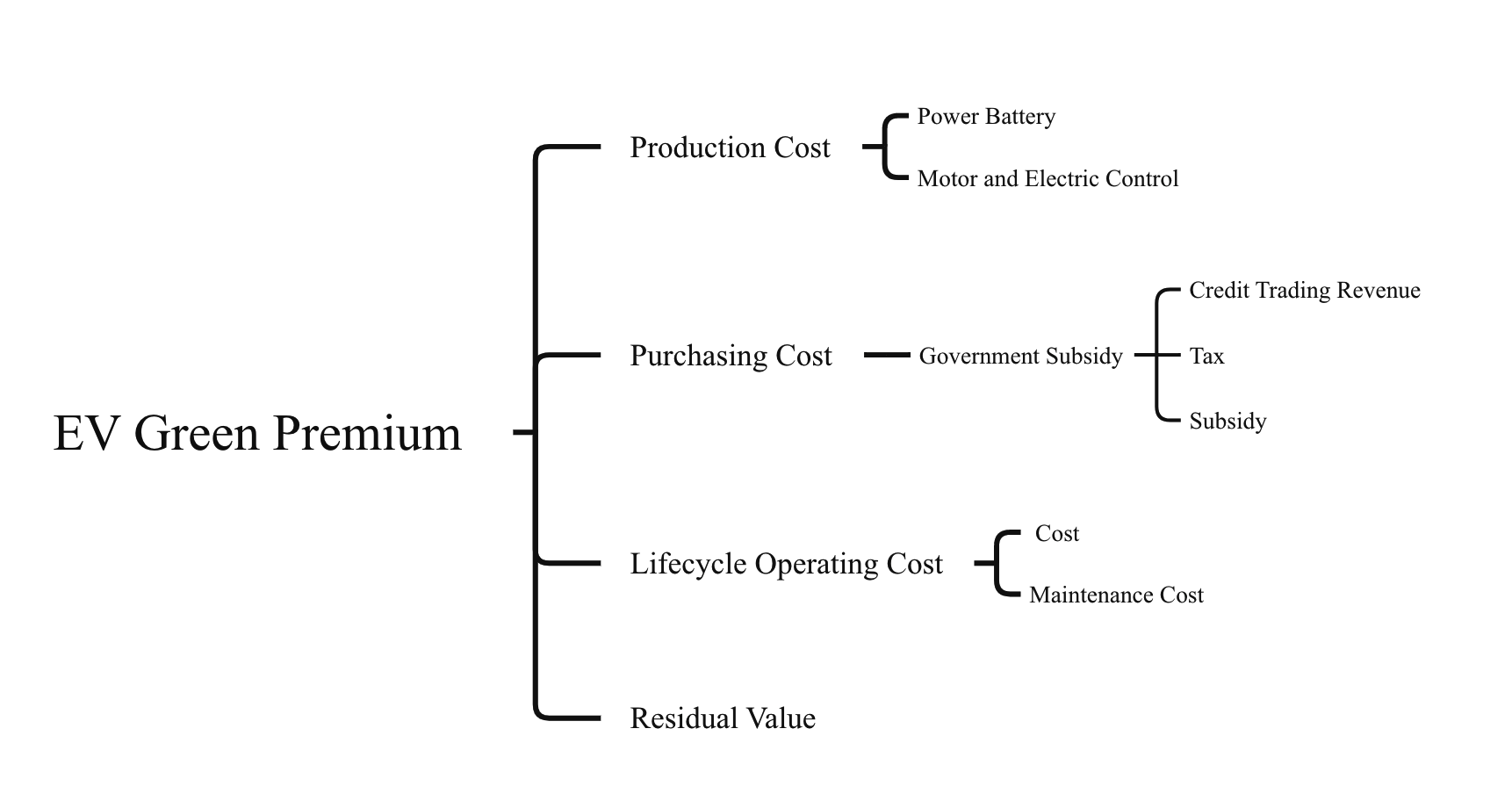}
    \caption{The Decomposition of Green Premium}
    \label{fig:my_label_1}
\end{figure}
For the purpose of the green premium calculation, all future costs are converted to net present value (NPV) at the time of vehicle purchase. Therefore, the NPV of the TCO is written as
\begin{equation}
\begin{aligned}
TCO_{NPV} = Production \: Cost + Government \: Subsidy \\ +\sum_{t=1}^{n}\frac{Annual \: Operating \:Cost}{(1+r)^t}-\frac{Residual \: Value}{(1+r)^n}
\end{aligned}
\end{equation}
\begin{equation}
\begin{aligned}
Government \: Subsidy= Credit \times Price_{credit} \\+ Tax +Subsidy
\end{aligned}
\end{equation}
\begin{equation}
\begin{aligned}
Annual \: Operating \: Cost = Fuel \: Cost + Maintenance \: Cost
\end{aligned}
\end{equation}
Government Subsidy includes Credit Trading Revenue, Tax and Subsidy. Residual Value is the end-of-life value and battery recycling cost of electric vehicles.

The samples in this paper come from the primary data of auto supplier research, and the Grade B-class sedan with the highest penetration rate in EV passenger cars is selected as the parameter benchmark. Tesla Model 3 (B-class sedan, 75kWh, 480km range) is the world's first and only million-selling electric car and is representative of electric car performance and price. Therefore, in this paper, a B-class fuel car with similar specifications and torque (75kWh battery approximately corresponds to a 2.0~2.4L supercharged engine) was selected as the control model and substituted into the model for data simulation.

\subsection{Production cost model: power cell system}
The main difference between EVs and ICEVs is the unique design and structure of the powertrain, which uses batteries, electric motors, and electric motor controllers instead of the internal combustion engine for propulsion. ICEV powertrain costs consist mainly of the engine, transmission, exhaust system, and ECU/sensors, while the EV powertrain system consists mainly of batteries, electric control/motors and high-voltage components, and other auxiliary components of comparable value volume to ICEVs. The battery, as the component with the highest unit price volume, accounts for about 35-40\% of the material cost and is considered to be an important obstacle for EV manufacturers to reduce the green premium. Therefore, this paper will begin with an analysis on the battery cost structure. 

\subsubsection{Battery cost}
Battery performance (range, energy density, etc.) and price are largely determined by the cathode material of the battery and its operational life cycle. This paper predicts the power battery cost based on two factors: Unit battery capacity (kWh) and unit battery material cost (RMB/kWh). When accounting for economics of different batteries vis-à-vis battery type, the lithium-ion battery pack is the current mainstream battery model, whereby nickel-cobalt-manganese ternary lithium-ion battery (NCM) accounts for more than 60\%, nickel-cobalt-aluminum ternary lithium-ion battery (NCA) accounting for 30\%, followed by lithium iron phosphate (LFP) and lithium manganate (LMO) accounting for the remainder of the battery market share. 

Lithium-ion batteries contain raw materials and packaging design materials, where the cost is comprised mainly of the following: positive and negative electrode materials (30\%), diaphragm (10\%), and electrolytes (5\%). In 2021, the power battery system price is about RMB 820/kWh. The cost of a single battery pack for long-range (mainly ternary lithium batteries) is about RMB 62,000, and short-range (mainly lithium iron phosphate batteries) is RMB 45,000. Compared to the United States and Europe, the absolute cost of the power system in the Chinese market is 20\% lower.

The battery cost model in this paper takes into account the current mainstream ternary lithium batteries (NMC and NCA) and lithium iron phosphate (LFP). The battery cost decline in the short and medium term has a strong certainty. From 2010 to 2021, the cost of ternary lithium batteries for electric vehicles dropped from RMB 7,500/kWh to RMB 820/kWh, a reduction of nearly 90\%. The cost reduction mainly derives from both innovative new materials, and the optimization of the battery structure. According to the industry market data, the average price of battery per kWh is expected to drop to 650 RMB/kWh by 2025. 

Due to the scale effects, the NMC battery prices are expected to fall rapidly. But this process will slow down in about ten years as raw materials will constitute a higher share of total battery price thereby setting a price floor on battery prices. For example, JRC forecasts that global demand for cobalt, a component in electric vehicle batteries (NMC811) in 2030 will reach about 80\% of total global cobalt mine production based on 2016 global cobalt production output. JRC indicates that in response to raw material shortages and soaring prices, the producers may need to shift to other battery material that is less dependent on cobalt. 

\subsubsection{Motor, electric control, and other high-voltage components}
Electric motors are to EVs is like engines are to ICEVs. Electric motors and electric control systems are key drive components accounting for 15\% of the material cost of EVs. Similar to the cost analysis of batteries, the sub-model simulates its two factors of unit battery capacity (kWh) and unit motor/electronic control cost. The downward cost pressure mainly derives from the improvement of the motor and electric control technology, which led to a significant decrease in material costs, and the introduction of automated production lines which led to a reduction in manufacturing costs. \par
In summary, the green premium model considering only material and manufacturing costs without considering government subsidies such as point trading is as follows.\par
Green premium model for manufacturing cost.\par
\begin{equation}
\begin{aligned}
\Delta p_{1} &= \frac{P_{EV}(t)-P_{ICEV}(t)}{P_{ICEV}(t)}\\
             &= \frac{(x \times y)+(p \times q)+z-p_{1}-p_{2}}{p_{1}+p_{2}}
\end{aligned}
\end{equation}
where $P_{EV}(t)$ is the material cost of an electric vehicle at moment t, $P_{ICEV} (t)$ is the material cost of an internal combustion engine vehicle at moment t, $x$, and $y$ are the unit battery material cost (yuan/kWh) and unit battery capacity (kWh), $p$ and $q$ are the motor/electric control material cost (yuan/kWh) and unit motor and motor power (kWh), $z$ is the cost of other high-voltage components, $P_1$ and $P_2$ are the engine and intake and exhaust system costs and automatic transmission costs, respectively.\par
In this paper, two B-class cars with similar specifications and torque are selected for comparison. The basic parameters of EV are based on Tesla's B-class electric sedan (75kWh, 480km range), which has the highest market penetration; the basic parameters of ICEV are based on B-class fuel cars (75kWh battery approximately corresponds to the 2.0-2.4L supercharged engine).\par
\subsection{Acquisition cost model: government subsidies}
During the acquisition phase, the Chinese government's incentive policies include a double credit trading policy, tax breaks, acquisition subsidy policy, and purchase restriction policy. Since the license plate restriction policy is only implemented in a few cities in China, it is not included in this model for the time being and is discussed separately below.
\subsubsection{Purchase tax}
The current vehicle purchase tax in China is 10\% and electric vehicles are exempt from the purchase tax until December 2022. In the case of the Tesla Model 3, for example, the 10\% purchase tax exemption equates to a credit of 16,000 RMB to 19,000 RMB for a single vehicle.
\subsubsection{Acquisition subsidy}
China's current purchase subsidies for EVs are entering a declining phase, decreasing year by year until 2022. Subsidies for vehicles depend on model, range, energy density, and vehicle energy efficiency, and this paper assumes that purchase subsidies will apply from 2020 to 2022, but not from 2023 to 2026. EVs priced below 300,000 RMB can apply for government subsidies according to specific conditions. Taking the Tesla Model 3 as an example, each vehicle can receive government subsidies of 12600 - 19800 RMB respectively.
\subsubsection{Double credit trading revenue}
The double credit policy is a key variable for vehicle producers to evaluate which power models to produce because it subsidizes manufacturers of EVs through a market-based mechanism. Whether the cost of credit trading is eventually passed on to consumers depends on the relative elasticity of supply and demand. In this paper, the impact of credits trading on the green premium is measured by assuming a fixed gross margin for manufacturers.\par

The trading orders of double credits in China exceed 2 million in 2020, and the total transaction amount exceeds 2 billion yuan. So this paper starts to calculate the point gain from 2020. According to the latest "Parallel Management Measures for Average Fuel Consumption of Passenger Vehicle Enterprises and Electrical Vehicle Credits" (2020) issued by MIIT, $CAFC$ points are the difference between the standard value and the actual value, denoted as $( T_{CAFC}-CAFC)$; positive credits of EV are the difference between the actual value and the standard value, denoted as $(NEV-T_NEV)$. positive credits of NEV are linked to $CAFC$ in a one-way, The trading price of the integral is denoted as $P_c$.\par

$CAFC$ credits are highly correlated with ICEV fuel consumption per vehicle, and NEV points are correlated with EV range, energy density, electric consumption intensity, and overall mass. Since positive CAFC credits cannot be traded, only the compliance cost of negative CAFC credits to fuel manufacturers and the trading revenue of positive NEV credits to EV manufacturers are calculated in this model. To simplify the calculation, it is assumed that the credits generated in the current year generate trading revenue in the current year (no carryover). A BEV can earn up to 3.4*1.5=5.1 NEV points and an ICEV (fuel consumption 6.49L/100km) generates about -0.11 CAFC credits as measured by the Chinese passenger vehicle fuel consumption management stage 5 (2021) standard.\par

$P_c$, as an exogenous variable, is determined by the supply and demand of points in the market in the current year and fluctuates greatly. The transaction price of double credits in the past 5 years ranged from as low as \$50/points to as high as \$3,000/points. If the industrial development does not match the policy goal, there may be a situation of oversupply or undersupply of positive points. 1:1 offset of negative CAFC credits by positive NEV points in one direction, based on the transaction price of RMB 2,000/credit in 2021, a single BEV can get up to RMB10,200,000 points compensation.\par

The acquisition cost green premium model is as follows.\par
\begin{equation}
\begin{aligned}
\Delta p_{2} &= \frac{P_{EV}(t)-Dev-S-(P_{ICEV}(t)+D_{IECV}+T_{ICEV}}{P_{ICEV}(t)+D_{IECV}+T_{ICEV}}\\
             &= \frac{\Delta M}{P_{ICEV}(t)+(NEV-T_{NEV})\times P_{c} +T}
\end{aligned}
\end{equation}

\begin{equation}
\begin{aligned}
\Delta M = P_{EV}(t)-(T_{CAFC}-CAFC)\times P_{c}\\-S-(P_{ICEV}(t)+(NEV-T_{NEV})\times P_{c} +T)
\end{aligned}
\end{equation}

\subsection{Full life cost model: cost in use and salvage value}
Based on the full life cycle theory, this section establishes a full life cycle cost model for electric and fuel vehicles by considering energy use, regular maintenance costs, and residual values of used vehicles over the lifetime of the vehicles. This paper argues that life-cycle cost is the key factor that affects whether EVs can go to the market. If the products themselves do not have cost competitiveness and performance advantages, it is difficult to achieve full industrialization with government subsidies alone.\par
\subsubsection{Fuel cost}
Fuel use is a key variable in the economic, environmental, and engineering analysis of motor vehicles, which directly affects the energy cost, driving range, and greenhouse gas emissions, and indirectly affects the residual value and life cycle cost of the vehicle. In this analysis, fuel used depends on fuel consumption and fuel price, and fuel consumption depends on driving cycle and energy consumption intensity, so the model calculates the energy consumption intensity (kWh/100km) and energy unit price  for EVs and ICEVs in a given driving cycle.\par

The energy consumption intensity is related to body weight, motor or engine efficiency, transmission efficiency, and body resistance. the fuel consumption efficiency of ICEVs is standardized according to the fuel consumption limit set by CAFC, with an average level of 6.43-6.7L/100km in 2021. the electric consumption intensity of EVs is based on the current industry reality and the average electric consumption standard for new pure electric passenger vehicles proposed by the State Council, which is currently about 13.00kWh/100km.\par

Regarding energy prices, the historical value of 92\# gasoline used in passenger cars is predicted in the basic model of this paper, and due to the volatility of oil costs, this paper assumes that the future floats around the average value of 7.5 RMB/liter. For the electricity price, based on the calculation for the historical household electricity price plus the charging post-service fee, it is calculated at RMB 1.2/kWh.
The annual mileage (VMT) determines the annual fuel and maintenance costs as well as the residual value of the used vehicle. For the basic model, this paper assumes 15,000 km per year (Center for Energy and Transportation Innovation, 2021 report) and a passenger car life of 10 years, assuming a full life cycle of 150,000 km.\par

\subsubsection{Maintenance costs}
Regular maintenance costs are a major source of operating costs for ICEVs, and one of the main differences from EVs. According to the owner's manual of each vehicle, ICEVs need to go back to 4s stores regularly for cleaning or replacement of oil, filter, air filter, air conditioning filter, spark plugs, brake pads, and other spare parts. The cost of regular maintenance is related to the mileage of the car, and according to the online auto repair cost estimation website, it needs more than 7,000 RMB per year for maintenance and repair costs.\par

EVs do not require high-priced maintenance items such as oil and three filters due to the fundamental difference between the fuel cars. EVs only need the cost of routine inspection and keeping them clean. In addition, in order to dispel consumers' doubts, the manufacturer proposes an 8-year or 150,000 km warranty for the entire vehicle, mainly referring to the warranty for the three electric systems (battery, motor, and electric control). The service life of lithium-ion batteries depends on their cycle number. The current service life of the lithium-ion battery is 3-5 years, and it needs to be replaced at least 2 times in the EV life cycle, and this cost is now borne by car manufacturers. Without considering the same basic maintenance costs as ICEV (such as tires), the EV with Tesla Model 3 as an example only requires an average annual maintenance cost of about 2,000 yuan, while the maintenance cost of EVs under 300,000 yuan is only 300 yuan/year.\par

\subsubsection{Residual value of used cars}
The depreciation method for fuel cars is relatively mature, and the residual value of a used car can usually be calculated based on the number of years of use and mileage.
The residual value of EVs is mainly determined by the battery pack life, and the price of battery packs is declining rapidly this year, so the residual value of used cars is usually calculated based on 70\% of the initial cost of the battery multiplied by the percentage of remaining battery life. However, compared to ICEVs, EVs have a shorter service life, and the three electric systems, which account for the largest proportion of the cost, lack residual value assessment standards, and due to the difficulty of pricing used cars and the lack of liquidity in the trading market, the remaining battery life of EVs and the residual value of used cars have greater uncertainty.
The life-cycle cost (TCO) model is composed of four parts: acquisition cost (i.e., purchase price, purchase tax, government subsidies, etc.), fuel cost, maintenance cost, and the residual value of used vehicles. The model assumes a discount rate of $r=5\%$ to evaluate the first purchase cost and the full life-cycle cost of ownership for fuel-fired vehicles and EVs and performs a sensitivity analysis of alternative discount rates.\par
The TCO for EVs and ICEVs of fuel vehicles can be expressed as\par
\begin{equation}
\begin{aligned}
&C_{EV}(t) = P_{EV} -D-S+\\
&\sum_{t=1}^{n}(\frac{VMT \times EC \times F_{e}(t) + M_{EV}(t)}{(1+r)^t})
-\frac{V_{EV}(t)}{(1+r)^n}
\end{aligned}
\end{equation}
Total Life Cycle Cost (TCO) Green Premium Model.
\begin{equation}
\begin{aligned}
&C_{ICEV}(t) = P_{ICEV} +D+T+\\
&\sum_{t=1}^{n}(\frac{VMT \times FC \times F_{g}(t) + M_{ICEV}(t)}{(1+r)^t})
-\frac{V_{ICEV}(t)}{(1+r)^n}
\end{aligned}
\end{equation}
where $C_{EV} (t)$ is the total cost of ownership of electric vehicles, $C_{ICEV} (t)$ is the total cost of ownership of internal combustion engine vehicles, $P_{EV}$ represents the market price of electric vehicles at time $t$, $P_{ICEV}$ represents the market price of internal combustion engine vehicles at time t, D represents the revenue (expenditure) from credit transactions, S represents the government acquisition subsidy, $T$ represents the taxes and fees associated with vehicle acquisition, n represents the life cycle of passenger vehicles, VMT (Vehicle Metres Travelled, per year) represents the average annual mileage of passenger vehicles, EC represents the electric consumption intensity (kWh/100km) of passenger vehicles, FC represents the electric consumption intensity (kWh/100km) of passenger vehicles, EC represents the electric consumption intensity (kWh/100km) of passenger vehicles. (Vehicle Metres Travelled, per year) represents the average annual mileage of passenger cars, EC represents the electric consumption intensity (Electricity consumption, kWh/100km) of electric vehicles, FC represents the fuel consumption intensity (Fuel consumption, L/100km) of fuel cars, F, L/100km), $F_e$ and $F_g$ represent the cost of electricity and gasoline, respectively, $M_{EV} (t)$ and $M_{ICEV} (t)$ represent the annual maintenance cost of EV and ICEV, respectively, and $V_{EV} (t)$ and $V_{ICEV} (t)$ represent the used vehicle residual value of EV and ICEV at time t, respectively.\par
\section{A generalized bass model considering the green premium}
\subsection{Classical generalized Bass model}
F.M. Bass developed the Bass model \cite{bass1969new} for predicting product sales, and the Bass diffusion model has two basic assumptions.
(a) Innovation coefficient: the parts of the product that are easily verifiable (e.g., price, model, policy, self-driving technology, etc.) are mainly influenced by mass media, called external influences. (b) Diffusion coefficient: the parts of product characteristics that are difficult to verify in the short term (e.g., reliability, durability, battery life, etc.) are mainly influenced by word-of-mouth communication from old users to potential users, called internal influence.\par

Based on these two assumptions users in the market are divided into innovators (geek users), who are mainly influenced externally, and imitators (followers), who are mainly influenced internally. The Bass diffusion model is the most widely used model to predict new product diffusion\cite{zhu2018forecasting}, which simulates an S-shaped new product penetration curve by constructing an innovation factor (external influence) and an imitation factor (internal influence). The expression of the Bass model is as follows.
\begin{equation}
\begin{aligned}
n(t) &= \frac{dN(t)}{d_t}\\
     &= (p \times (M-N(t))+q\times \frac{N(t)}{M} \times (M-N(t))
\end{aligned}
\end{equation}
In equation (9), $n(t)$ is the number of new product adopters at moment $t$; $p$ is the innovation coefficient (spontaneous adoption rate) and $q$ is the imitation coefficient (imitation adoption rate); $M$ is the market potential over the product life cycle and $N(t)$ is the cumulative number of adopters in the total user base up to t (excluding n(t)); in 1994 Bass added price $x(t)$ as a new decision variable in his article. IBM, Kodak, and Hewlett - Packard have used this model to successfully predict the diffusion of new products\cite{renbin@2013diffusionmodel}.\par
The generalized Bass diffusion model is expressed as
\begin{equation}
\begin{aligned}
n(t) &= \frac{dN(t)}{d_t}\\
     &= (p \times (M-N(t))+q\times \frac{N(t)}{M} \times (M-N(t)) \times x(t)
\end{aligned}
\end{equation}
The $x(t)$ in equation (10) is the decision influence coefficient of potential adopters at moment t. It is worth noting that innovators are not affected by the time of adoption, while imitators depend on social interactions with these innovators. Initially only $p$ plays a role, then once innovation adoption begins, $q$ takes over the diffusion effect, and diffusion slows down as $N(t)$ keeps approaching $M$ according to the research by \cite{kumar2022comparative}.
\subsection{A generalized bass model considering the green premium}
By rationalizing x(t) in the generalized Bass model, the impact of the incoming green premium on the electric vehicle sales market can be considered. For the EV market in China, the following factors are incorporated into the Bass model for correction in this paper.\par

(1) Economic cost and government subsidies. Most of the current passenger car consumers consider the tool attributes of EVs and purchase EVs not for environmental protection, but for cheaper comprehensive life-cycle costs than a gasoline vehicle. So the total cost of ownership (TCO) is included in the Bass model as an important reference to influence consumer decisions. For the generalized government subsidies, including the purchase subsidies introduced by the Chinese government, the purchase tax exemptions, the income from credit trading, and the "shadow value" brought by the EV license plate, combined with the time dimension to measure the impact after the subsidy withdrawal.\par

(2) Technological innovation and convenience of experience. The ultimate marketization of EVs does not only depend on government subsidies and competitive market prices, but also on the improvement of product performance, which is reflected in the battery capacity and range, the intensity of electric consumption, and the convenience of charging. In this paper, we mainly consider the impact of fuel economy on the conversion of potential EV consumers due to the intensity of electric consumption per vehicle.\par

(3) Social climate and market acceptance. The maturity of technology on the supply side alone does not mean that consumers will choose EVs. Policies and major manufacturers are needed to guide the demand side and drive the market to change consumption habits, in line with the basic assumption of diffusion of innovative behavior in the Bass model. Take Tesla as an example, in the early stage, it targeted the high-income group with environmental consciousness and the "geek users" of social celebrities (such as Silicon Valley entrepreneurs and Hollywood stars) to build the luxury models Roadster and Model S sports cars. After that, they commercialized electric cars by developing the high-end sports car market, and then developed sedans and non-luxury  to enter the broader market of "followers".\par

The formula can be expressed as follows.
\begin{equation}
\begin{aligned}
n(t) = (p \times (M-N(t))+q\times \frac{N(t)}{m} \times (m-N(t)) \times x(t)
\end{aligned}
\end{equation}
In equation (11), $n(t)$ is the number of newly adopted electric vehicles (not the cumulative number of adopters) at moment t, m is the maximum market potential, $N(t)$ is the cumulative number of adopted electric vehicles as of moment $t$, $p$ is the internal influence (innovation coefficient), and $q$ is the external influence (imitation coefficient).
\begin{equation}
\begin{aligned}
x(t) &= 1 + \Delta p_{3} \times \beta 
     &= 1 + \frac{C_{EV}(t)-C_{ICEV}(t)}{C_{ICEV}(t)}\times \beta
\end{aligned}
\end{equation}
In equation (12), $x(t)$ is the potential adopter decision influence coefficient at moment $t$. In this paper, we assume that the purchase decision of electric vehicles is mainly influenced by the relative cost of old and new technology, $C_{EV} (t)$ is the total cost of ownership of electric vehicles, $C_{ICEV}(t)$ is the total cost of ownership of internal combustion engine vehicles, $\Delta p_3 = (C_{EV} (t) - C_{ICEV}(t))/(C_{ICEV} (t))$ is the green premium of electric vehicles at moment $t$, $\beta$ is the price influence coefficient, and the calculation of $C_{EV} (t)$ and $C_{ICEV}(t)$ can be found in equations (6) and (7).

\subsection{Bass model parameter fitting}
Standard parameter estimation methods for Bass models include ordinary least squares (OLS), maximum likelihood (MLE), and sequential search-based nonlinear least squares (SSB - NLS). With the development of computing technology and optimization algorithms, Genetic Algorithm (GA) \citep{katoch2021review}, Ant Colony Algorithm \citep{dorigo2006ant}, Simulated Annealing Algorithm \citep{van1987simulated}, etc. are also used for parameter estimation of the Bass model and can obtain more accurate prediction results. In the case where there are fewer data points, the parameter space is multi-peaked and the model is essentially nonlinear, the genetic algorithm has a greater possibility of converging to the global optimum and the prediction results are more accurate. Therefore, this paper uses nonlinear least squares with genetic algorithms for parameter estimation.

\section{Case study}
\subsection{Algorithm test conditions and parameter descriptions}
In order to ensure the reliability of the conclusions, this paper sets certain values or value domains for the above parameters. The key assumptions to be calculated are described in detail in Table 1.\par

\begin{table}
    \fontfamily{ptm}\selectfont
	\caption{Relevant variables of the full life cycle cost model}
    \begin{tabular}{llll}
   
   \toprule 
       & Symbol  &  Variable & Value(Range) \\ 
   \midrule 
   \multirow{7}{*}{Production Cost}& x (RMB/kWh) & Battery material cost & 820\\
   \multirow{7}{*}{}& $y$ (kWh) & Battery capacity & 75\\
   \multirow{7}{*}{}& $m_{1}$ (RMB/kWh) & Motor material cost & 65\\
   \multirow{7}{*}{}& $m_{2}$ (kWh) & Motor system capacity & 200\\
   \multirow{7}{*}{}& $m_{3}$ (RMB) & Other high voltage component cost & 6000\\
   \multirow{7}{*}{}& $m_{4}$ (RMB) & Intake and exhaust system costs & 16000\\
   \multirow{7}{*}{}& $m_{3}$ (RMB) & Automatic transmission costs & 11000\\ \hline 
   \multirow{8}{*}{Purchase Cost Model}& $T$(\%) & Purchase tax (EV exempt) & 10 \\
   \multirow{8}{*}{}& $S$ (RMB) & Subsidy(for EV) & [0,18000]\\
   \multirow{8}{*}{}& $D$ & Credit trading revenue & $[3.8,4.6]\times p_{c}$\\
   \multirow{8}{*}{}& $p_{c}$ & NEV trading price & $[1000,2000]$\\
   \multirow{8}{*}{}& $CAFC$ & Actual value of fuel consumption(ICEV) & $[6.5,7.5]$\\
   \multirow{8}{*}{}& $T_{CAFC}$ & Threshold value of fuel consumption(ICEV) & $[4,10]$\\
   \multirow{8}{*}{}& $NEV$ & Actual Value of EV Credits & $[1.6,6]$\\
   \multirow{8}{*}{}& $T_{NEV}$ & Threshold Value of EV Credits & 0\\ \hline
   \multirow{8}{*}{}& $n$ (year) & Life cycle of passenger cars & 10\\
   \multirow{8}{*}{}& $VMT$ (km) & Average annual passenger vehicle miles & 15000\\ 
   \multirow{8}{*}{Driving Cost Model}& $EC$ (kWh/100km) & Electricity Cost of EV & 13 \\
   \multirow{8}{*}{}& $FC$ (L/100km) & Gasoline cost of ICEV & 8.5\\
   \multirow{8}{*}{}& $F_{e}(t)$ (RMB/kWh) & Electricity price(including charging fee) & 1.2\\
   \multirow{8}{*}{}& $F_{g}(t)$ (RMB/L) & Gasoline price & 7.5\\
   \multirow{8}{*}{}& $M_{EV}(t)$ (RMB/year) & Maintenance cost of EV & 7000  \\
   \multirow{8}{*}{}& $M_{ICEV}$ (RMB/year) & Maintenance cost of ICEV & 2000 \\ \hline
   \multirow{3}{*}{Residual Value}& $V_{EV}(t)$ (RMB) & Residual value of EV & [30000,40000] \\
   \multirow{3}{*}{} & $V_{ICEV}(t)$ (RMB) & Residual value of ICEV & [60000,70000]\\
   \multirow{3}{*}{} & $r$ (\%) & Discount rate & 5\\

\bottomrule 
\end{tabular}
\end{table}

For the battery cost, from 2010 to 2021, the cost of ternary lithium batteries for electric vehicles has fallen from RMB 7,500/kWh to RMB 820/kWh. It is a reduction of nearly 90\%, with the cost reduction mainly coming from the evolution of the material system and optimization of the battery structure. According to the industry market data, the average price of battery per kWh is expected to reach 650 RMB/kWh by 2025. Besides the model uses the 2021 Fuel Consumption Limits for Passenger Cars to set a target of 4.0L/100km (NEDC conditions) for fuel vehicles by 2025, with $T_{CAFC}$ values currently set in the 4-10L/100km range. The credits trading price varies a lot from within 1000 RMB/points in 2018, about 500 RMB/points in 2019, up to about 1000-3000 RMB/points in 2020, and back down to 2000 RMB/points in 2021. The uncertainty of the points price is high. So this paper establishes a larger range of 1000-2000 RMB/points. The higher credits trading price will ensure that the policy effect of the model is more obvious and does not affect the accuracy of the core findings.

\subsection{Green premium model output results}
According to the constructed green premium model, key technical parameters of the current technical level of new energy vehicles are inputted for simulation. In this paper, the B-class electric sedan (75kWh, 480km range) with the highest market promotion degree is used as the reference standard for EV models, and the corresponding B-class fuel car (2.4L supercharged engine) is used as the reference standard for internal combustion engine car models, and the data simulation operation is conducted under the conditions of EV subsidy policy and usage cost in 2021. It is found that the production cost of EV (B-class long-range model) shows a higher green premium (44\%), but the full life-cycle cost has been discounted (-15\%).\par
\begin{figure}[ht]
    \centering
    \includegraphics[width=1\textwidth]{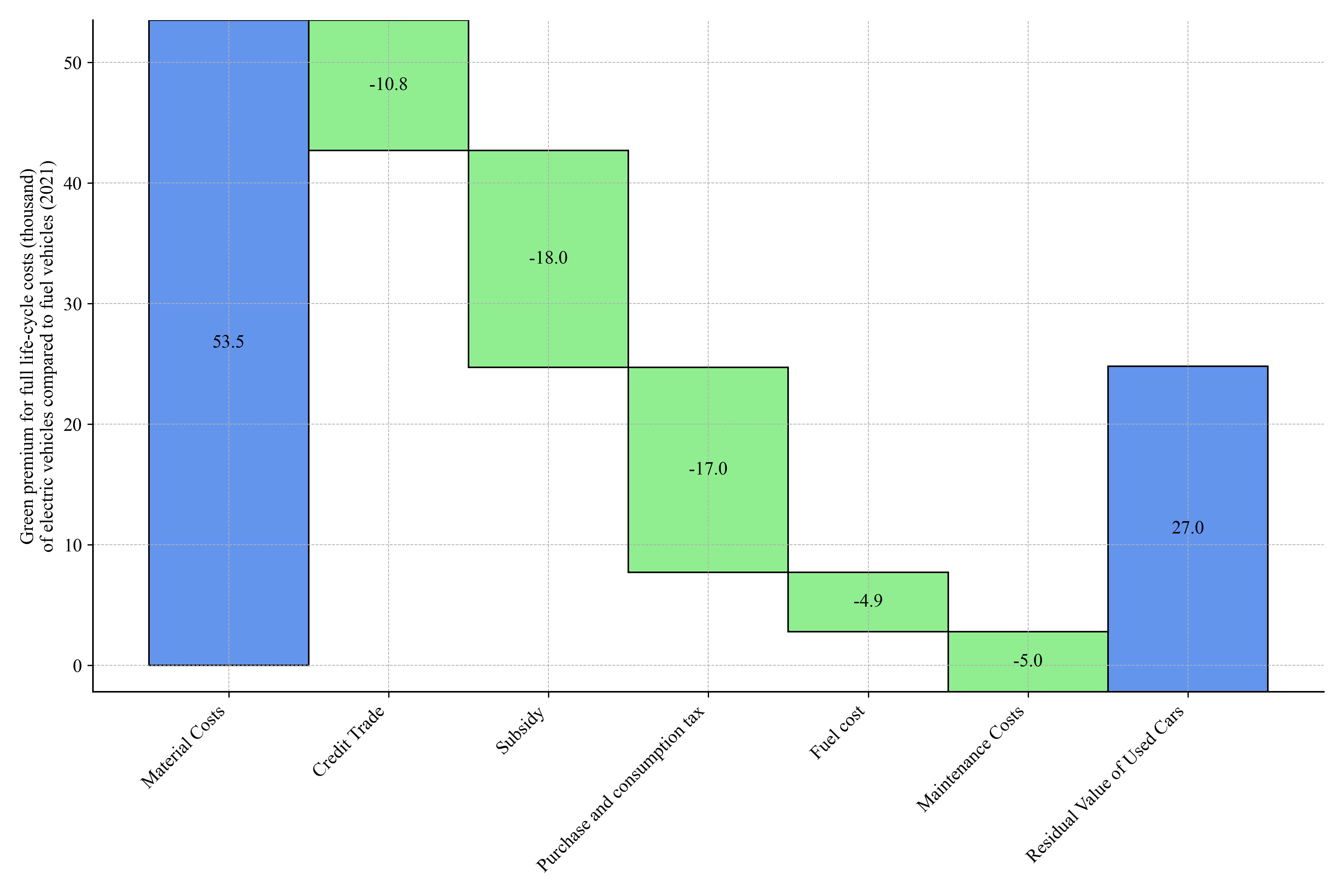}
    \caption{Green Premium for Full Life Cycle Cost of Electric Vehicles Compared to Fuel Vehicles (2021)}
    \label{fig:my_label_2}
\end{figure}
The output of the model is shown in Figure 2. The production cost of EV has a 44\% green premium compared to ICEV, indicating that the EV manufacturers have weak or even negative profitability of their own in the early stage of development. This is mainly due to the high cost of batteries, motors and electric control systems, and other high-voltage components at the early stage of industrialization when EV technology and manufacturing processes are still immature. From the perspective of pure material cost, the green premium of EV production cost (44\%) is close to its battery cost (37\%), so the reduction of battery cost and the growth of production scale will be the key to the downward green premium of EV material cost.
In the acquisition cost, policy subsidies (including credit trading, purchase tax credits, and acquisition subsidies) make up 36\% of the premium, and there is only an 8\% green premium in the actual consumer acquisition cost for EVs. In this paper, by constructing government subsidies as exogenous variables in the model to influence the marginal profit of enterprises, traditional automakers are constrained by the compliance cost under the mandatory regulations of technology on the one hand, and the additional revenue brought by policy subsidies such as double credits trading, acquisition subsidies and tax incentives for EVs on the other hand, which greatly reduces the acquisition threshold of EVs and also argues the subsidy policy from the perspective of cost-effectiveness. The effectiveness of the subsidy policy is also proven from a cost-benefit perspective.
In the cost of use, EVs reduce the green premium by 22\% through the economy of new energy, and the longer the use of EVs, the more obvious the operating cost advantage. The battery cost of a pure EV is about 60-70\% lower than the fuel cost of a fuel car, and the maintenance cost is 70\% lower due to the simpler structure of a pure EV. Compared to ICEVs, using EVs can save about \$10,000 per year in fuel and maintenance costs, which means the total cost of using EVs for 1.6 years is equal to that of ICEVs.
The used vehicle residual value is also considered in the full lifecycle cost, and the cost per kilometer of use after lifecycle amortization is \$1.52/km for EVs and \$1.80/km for ICEVs, achieving a -15\% green discount. This suggests that although fuel cars may appear cheaper in the early stages because of the lower purchase price, pure EVs have more of an operating cost advantage in the use phase, and the longer the car is driven the more cost-effective it is to use EVs.
\subsection{Dynamic prediction of the green premium for electric vehicles}
Based on the full life cycle cost green premium model constructed above, this paper analyzes and compares the economic competitiveness between different vehicle technology based on the assumptions of key factors such as acquisition costs, taxes, and subsidies, fuel costs, maintenance costs, and simulates and predicts the transition process of electrification in the automotive market. In addition, a sensitivity analysis is conducted to identify the key factors affecting the market penetration of electric vehicles.
The long-range electric vehicle is compared with the standard B-class pure electric vehicle with a 75kWh battery pack and NEDC range of 668km, compared to the equivalent model of fuel cars, and the green premium model output is shown in Figure 3 below.
\begin{figure}[ht]
    \centering
    \includegraphics[width=0.9\textwidth]{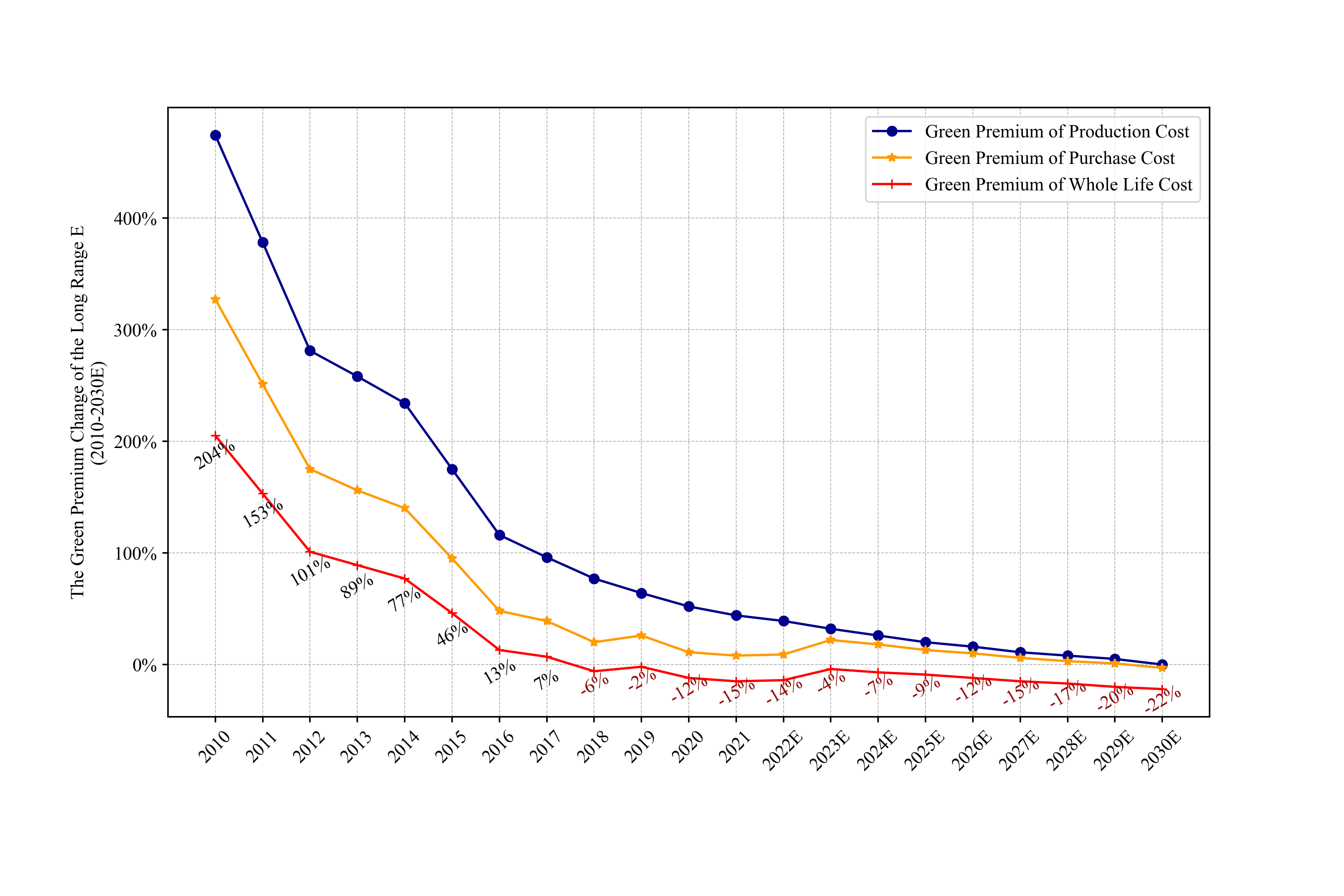}
    \caption{Green Premium Change for Long-Range Electric Vehicles (2010-2030E)}
    \label{fig:my_label_3}
\end{figure}

The short-range electric vehicle with a 60 kWh battery pack and an NEDC range of 455 km as standard for a B-class pure electric vehicle is compared to the equivalent model of a fuel vehicle, and the green premium model output is shown in Figure 4 below.
\begin{figure}[ht]
    \centering
    \includegraphics[width=0.9\textwidth]{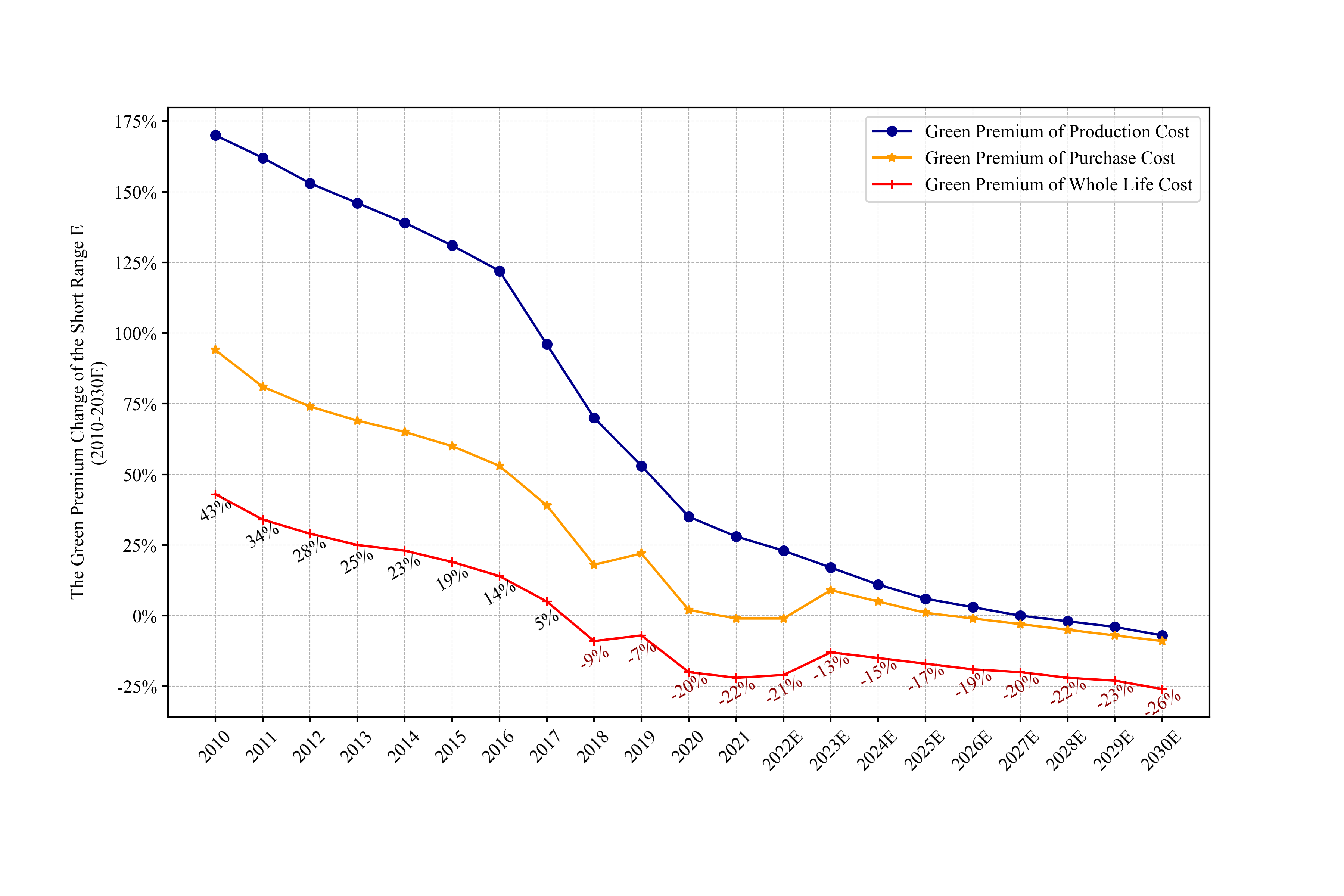}
    \caption{Green Premium Forecast for Short-Range Electric Vehicles (2021-2030)}
    \label{fig:my_label_4}
\end{figure}
In this paper, we will judge the speed of technology market adoption by the time when the green premium drops to zero, which is the parity point between EVs and ICEVs. Since the difference in overall mass and price between EVs and equivalent ICEVs mainly comes from their powertrains, and EVs with longer range tend to be heavier, the green premium for long-range EVs will be higher, so short-range EVs will achieve parity sooner.
As shown in Figure 4, short-range EVs (300-500km) have already achieved a green discount on full life-cycle costs in 2018, i.e., EVs cost less than fuel cars, and acquisition costs will rebound in 2022 after the subsidy rollback but will achieve parity with fuel cars in 2025-2026, and eventually achieve parity in production costs in 2027-2028. As shown in Figure 3, long-range EVs (500km+) also achieve a green discount on full life-cycle costs in 2018, but their production costs are higher compared to short-range, so parity between acquisition and production costs is not expected to be achieved until 2030.
The order of cost parity for both models is "lifecycle cost → acquisition cost → production cost," as the economics of EV use drive the full lifecycle cost toward parity. Before the green premium for total cost drops to zero, EVs already bring consumers significant economic savings from a cost-of-use perspective, saving about 60\%-70\% on fuel and maintenance costs compared to fuel vehicles, with an absolute value of 7,000-10,000 RMB/year. When the reduction in usage costs brought about by EVs is greater than the premium in upfront acquisition costs, the point of parity in full life-cycle costs is reached.\par
It is worth noting that there is an overall increase in the green premium in 2023, which is mainly due to the completion of the subsidy policy rollback in 2022 and the discontinuation of acquisition subsidies and purchases tax reductions for EVs from 2023 onwards, resulting in a rebound in acquisition costs. Therefore, the decline in EV green premium before 2023 is mainly driven by policy subsidies and the economics of their cost of use. The decline in green premium after 2023 mainly comes from the significant decline in EV production costs.\par
\begin{table}
\fontfamily{ptm}\selectfont
	\caption{Change in levelized cost LCOD (Yuan/km)}
    \begin{tabular}{llll}
   \toprule 
    Year & ICEV  &  EV(Long Range) &EV(Short Range) \\ 
   \midrule 
    2010 & 1.87 & 5.71 &2.68\\
    2011 & 1.87 & 4.73 &2.50\\
    2012 & 1.86 & 3.74 &2.40\\
    2013 & 1.85 & 3.49 &2.32\\
    2014 & 1.84 & 3.26 &2.26\\
    2015 & 1.84 & 2.69 &2.19\\
    2016 & 1.83 & 2.07 &2.09\\
    2017 & 1.82 & 1.95 &1.92\\
    2018 & 1.81 & 1.70 &1.65\\
    2019 & 1.80 & 1.77 &1.68\\
    2020 & 1.81 & 1.59 &1.45\\
    2021 & 1.80 & 1.52 &1.41\\
    2022E & 1.80 & 1.54 &1.41\\
    2023E & 1.79 & 1.72 &1.56\\
    2024E & 1.79 & 1.68 &1.52\\
    2025E & 1.80 & 1.6 &1.49\\
    2026E & 1.80 & 1.59 &1.47\\
    2027E & 1.82 & 1.55 &1.45\\
    2028E & 1.83 & 1.51 &1.43\\
    2029E & 1.83 & 1.47 &1.41\\
    2030E & 1.85 & 1.45 & 1.37\\
\bottomrule 
\end{tabular}
\end{table}

\subsection{Sensitivity analysis}
Based on the base model, a sensitivity analysis was conducted on the key parameters affecting the green premium of electric vehicles, as shown in Table 3, including the assumed battery materials and properties, government subsidy status, fuel cost, fuel consumption efficiency, maintenance cost, driving range and the residual value of used vehicles. The sensitivity coefficient is the ratio of the rate of change of the green premium to the rate of change of the impact factor.

\begin{table}
\fontfamily{ptm}\selectfont
	\caption{Sensitivity Analysis}
    \begin{tabular}{llllll}
   \hline    
    Factor & \multicolumn{4}{c}{Range}  &  Sensi\\ 
   \hline 
         & -20.0\% & -10.0\% &10.0\% &20.0\% &\\
    \textbf{Production} &   &   &  &  & \\
    Battery (800) & -36.0\% & -18.0\% &18.0\% &36.0\% &180\%\\
    Battery (650) & -21.9\% & -10.9\% &10.9\% &21.9\% &110\%\\
    Battery (500) & -14.9\% & -7.5\% &7.5\% &14.9\% &75\%\\
    \textbf{Subsidy}&   &   &  &  & \\
    Credit & 5.0\% & 2.5\% &-2.5\% &-5.0\% &-25\%\\
    Tax Rate(10\%) & -39.3\% & -19.6\% &19.6\% &39.3\% &197\%\\
    Subsidy  & 16.3\% & 8.2\% &-8.2\%  &-16.3\%  &-82\% \\
    \textbf{Cost} &  &  & & &\\
    Oil Cost(6L) & 21.8\% & 11.1\% &-11.1\% &-21.8\% &-109\%\\
    Oil Cost(4L) & 27.9\% & 14.1\% &-14.1\%  &-27.9\%  &-140\% \\
    Elec Cost(13kWh) & -8.3\% & -4.1\% &4.1\% &8.3\% &42\%\\
    Range(15000) & 16.7\% & 9.1\% &-9.1\% &-16.7\% &84\%\\
    Elec Price(1.2) & -1.5\% & -0.8\% &0.8\% &1.5\% &8\%\\
    Oil price(7.5) & 20.8\% & 10.6\% &-10.6\% &-20.8\% &-104\%\\
    \textbf{Residual} &  &   &  &  & \\
    EV Residual & 6.9\% & 2.8\% &-2.8\% &-6.9\% &-34\%\\
    Discount Rate & -0.9\% & -0.5\% &0.5\% &0.9\% &5\%\\
    
\hline  
\end{tabular}
\end{table}
Comparing the absolute values of sensitivity coefficients, it is found that battery unit price, purchase tax rate, fuel consumption, and gasoline price have the greatest impact on the green premium of EV full life cycle cost, with the absolute value of sensitivity coefficients exceeding 100
Production cost is the core factor affecting the green premium of EVs, which is influenced by the cost of the power battery system. For battery material prices are divided into three scenarios measured, BASE scenarios are 800, 650, and 500 RMB/kWh, representing the projected levels in 2021, 2025, and 2030. As the unit cost of the battery system decreases (from \$800/kWh to \$500/kWh), the percentage of the overall vehicle cost decreases, and the sensitivity factor decreases, while the rate of price reduction is slowed down year by year by the battery performance approaching the theoretical value and the limited amount of raw material reserves.
Another key variable affecting the cost competitiveness of EVs is the energy consumption intensity, i.e., the fuel consumption efficiency of ICEVs versus the electric consumption efficiency of EVs. The sensitivity of fuel consumption and fuel price is higher than electricity consumption and electricity cost, which is in line with the actual user experience. Since gasoline prices are more costly relative to electricity, the fuel cost of a fuel car is three times higher than that of an EV for 100 km, and each unit change in fuel consumption intensity brings a more significant change in usage cost, while the electrical energy used by EVs is inherently significantly more economical, so consumers are not as sensitive to changes in electrical consumption intensity.
\subsection{Genetic algorithm-based parameter estimation}
\subsubsection{Estimation of the maximum market potential $m$}
The maximum market potential depends on the maturity of the product, the coverage of public supporting infrastructure, and the strength of government policies. According to the development target of 20\% penetration of new energy vehicles by 2025 proposed in the State Council's Electrical Vehicle Industry Development Plan (2021-2035) (2020) and the Chinese Society of Automotive Engineering's 20\% by 2025, 40\% by 2030 and 50\% by 2035 proposed in the Energy Conservation and Electrical Vehicle Technology Roadmap 2.0 (2020) Research-based policy goals, this paper assumes that the largest market for electric vehicles in 2030 is 40\% of overall vehicle ownership. According to the calculation of 302 million vehicles in China in 2021, vehicle ownership in China will reach 640 million in 2030, and the penetration rate of new energy vehicles will be 40\%, 80\% of which will be pure electric vehicles (EV), which will result in a maximum market potential of about 21 million vehicles.
\subsubsection{Estimation of innovation coefficient $p$, imitation coefficient $q$, and price impact coefficient $\beta$}
In this model, the objective function of parameter fitting includes three items: (1) the minimum variance of deviation between predicted and actual values; (2) there are only 12 historical data points of electric vehicles, and the small database in the early period and the big influence by policies are easy to cause data distortion, so the calculation weight of deviation of 4 data points after 2018 is increased; (3) considering the low value of annual sales in the early period of electric vehicle development, the first 5-year value is less than 100,000 units, the model prediction result is easily less than zero, which is not in line with the actual situation, so the constraint of making the prediction value greater than zero needs to be added. The parameters of $p$, $q$, and $\beta$ are fitted by solving for the minimum value of the objective function.\par

The program for estimating the parameters of the Bass model by the genetic algorithm was written through Matlab with the following rule settings: population size of 800, crossover probability of 0.8, variance probability of 0.1, and the maximum number of iterations of 500. The original data from 2010-2021 were input into the model with m=21000 and the output results were as follows.\par

The parameter fitting results are consistent with the empirical values of historical studies and p<q, indicating that EVs are successful innovation diffusion products. $\beta$ is negative, indicating that as the green premium of EVs compared to fuel cars decreases, the annual sales of EVs rise and the absolute value is large, indicating a higher sensitivity to price for potential adopters.\par

Based on the parameter estimation results, the following expressions are obtained.\par

The core assumptions in this expression are built into the green premium in Chapter 3, i.e., the forecasted values of the $\Delta p_{3}$ EV lifecycle green premium for 2022-2030 in Chapter 3 are brought into the model to obtain the forecasted electric vehicle sales in China for 2022-2030.
\begin{figure}[ht]
    \centering
    \includegraphics[width=0.8\textwidth]{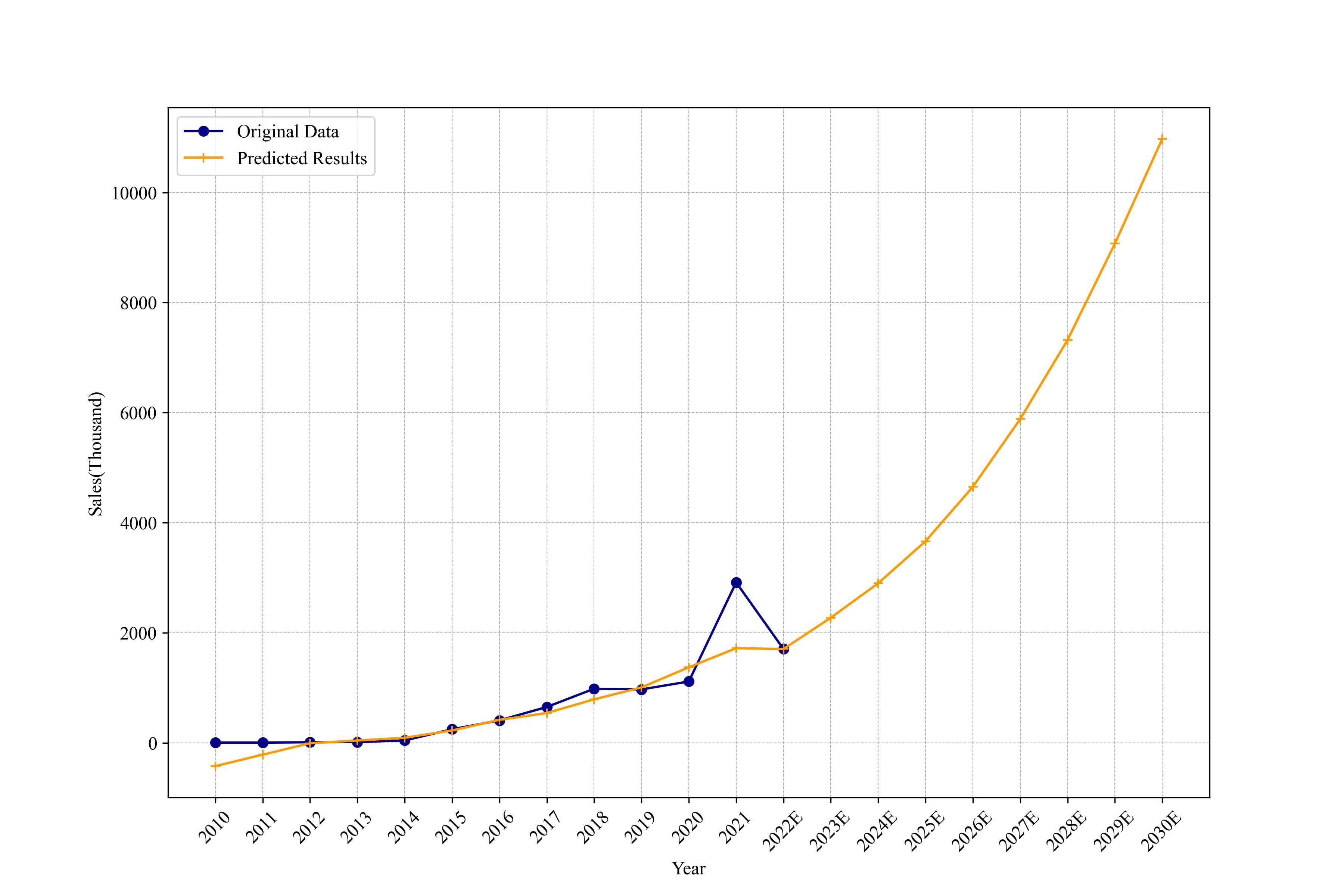}
    \caption{China Pure Electric Vehicle Sales Forecast (2010-2030)}
    \label{fig:my_label_5}
\end{figure}
As shown in Figure 5, electric vehicles are the more successful innovation diffusion product. With the natural scenario of existing policy incentives and technological innovation, the full life-cycle green premium for pure EVs will reach -22\% in 2030, i.e., the full life-cycle cost is 22\% lower than the cost of fuel vehicles, and the annual sales of pure EVs are expected to reach 10.97 million units and 54.41 million units cumulatively in 2030. In addition, although the annual sales of EVs are on an upward trend, they are not rising year by year and may experience sales fluctuations or even a brief decline during the painful period of subsidy withdrawal (e.g., 2022).
\subsubsection{The effectiveness of the generalized Bass model}
We also use the vanilla Bass model to fit the sales and compare the generalized Bass model and the traditional vanilla Bass model. The results are as follows.
\begin{figure}[ht]
    \centering
    \includegraphics[width=0.8\textwidth]{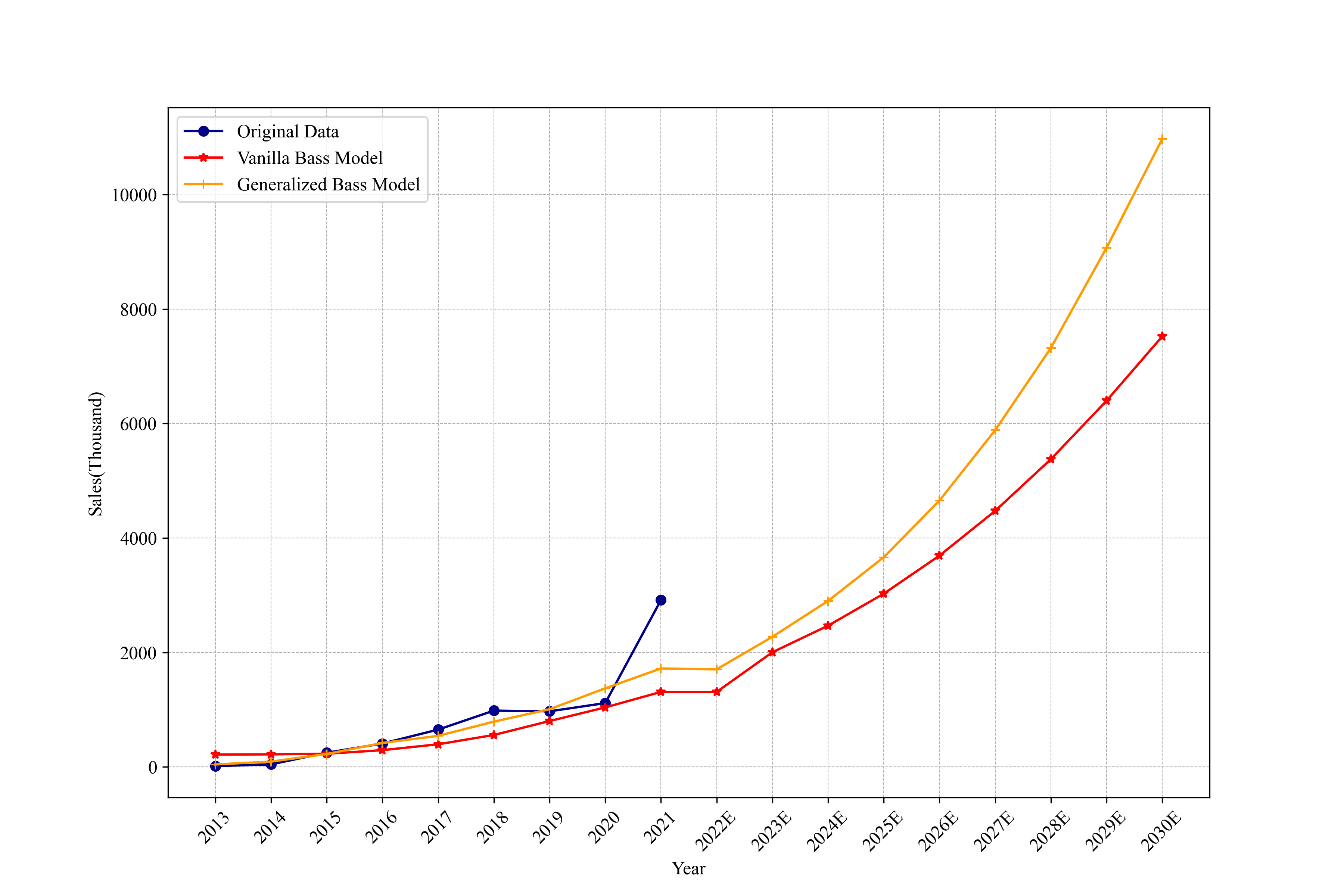}
    \caption{The Comparison of the Generalized Bass Model and Vanilla Model}
    \label{fig:my_label_6}
\end{figure}
The $R^2$ of the traditional vanilla Bass model is 0.6 and the Generalized Bass model is 0.77 which proves the effectiveness of the Generalized Bass model. Besides, we can also from the figure that the predicted results of the Generalized Bass model are more closed to the original data.
\section{Conclusion}
This paper combines quantitative and qualitative research methods to construct a multi-factor EV green premium model, which simulates the green premium decline process from 2010-2030 based on primary data from case studies. The following conclusions can be reached:

(1) Green premium is a key variable in realizing electrification of automobiles

This paper quantifies the factors affecting green transition by constructing a full life-cycle green premium model for ICEVs vis-à-vis EVs. The green premium decline in China's automotive industry from 2010 to 2030 is simulated using primary data collected from new energy vehicle manufacturers. The current full life-cycle cost of electric vehicles has achieved a green discount of -15\%, and policy subsidies and technological innovation are the core factors of the declining green premium. Under the pushback of technology-mandated policies, traditional automakers raise production costs to improve ICEV fuel economy, while EVs bring down production costs through a combination of lower battery costs, lower electric consumption intensity, and improved battery life through technological advances. The results show that the green premium of EVs will continue to decline while policy subsidies decrease, and will gradually achieve parity with ICEVs in the order of "life-cycle cost → acquisition cost → production cost" in the future. Short-range EVs will be the first to achieve parity in acquisition cost in 2025-2026, and long-range EVs will achieve parity around 2030.\par

(2) Predicting the penetration curve of electric vehicles through the green premium

In this paper, we modify the Bass diffusion model based on the green premium, select the time series data of EV diffusion in the Chinese market from 2010-2021, fit the parameters with a genetic algorithm, and forecast the EV market penetration in the next 10 years under different scenarios.
It is found that the modified Bass model performs better compares with the classical prediction models.
The EV market penetration rate is highly sensitive to the green premium. The results show that EVs are successful innovation diffusion products with high price sensitivity, and the market penetration rate is largely dependent on their green premium. The forecast results show that China's EV market may experience a slowdown or even a decline in growth in the short term, and will maintain high growth in the medium to long term, with annual sales expected to reach 10.97 million units and cumulative sales of 53.66 million units by 2030, with a penetration rate of about 39\%. Based on the scenario analysis, it is found that the double credits policy and license restriction policy plays a key role in the high growth rate of new energy vehicle penetration, and eventually through technological innovation and product scaling make EVs truly competitive in the market, from innovation to a larger scale proliferation stage.

(3) Green Premium brings a new analysis framework for the global carbon-neutral path from the perspective of low carbon technology industrialization

As a general analysis framework, the green premium can answer which industries have a high urgency for transformation and which technology are highly feasible for industrialization, and further predict the diffusion rate of different technology or products. Although the green premium varies widely across economic sectors due to factors such as the maturity of alternative technology, cost reduction potential, and compatibility of application scenarios, this paper uses the electrification transition in the automotive industry as an example to explain how the green premium predicts the industrialization of low-carbon technology and can be replicated in similar industries such as power, building materials, chemicals, and steel in the future.

\printcredits

\bibliographystyle{cas-model2-names}

\bibliography{cas-refs}

\begin{thebibliography}{23}
\expandafter\ifx\csname natexlab\endcsname\relax\def\natexlab#1{#1}\fi
\providecommand{\url}[1]{\texttt{#1}}
\providecommand{\href}[2]{#2}
\providecommand{\path}[1]{#1}
\providecommand{\DOIprefix}{doi:}
\providecommand{\ArXivprefix}{arXiv:}
\providecommand{\URLprefix}{URL: }
\providecommand{\Pubmedprefix}{pmid:}
\providecommand{\doi}[1]{\href{http://dx.doi.org/#1}{\path{#1}}}
\providecommand{\Pubmed}[1]{\href{pmid:#1}{\path{#1}}}
\providecommand{\bibinfo}[2]{#2}
\ifx\xfnm\relax \def\xfnm[#1]{\unskip,\space#1}\fi
\bibitem[{Adepetu and Keshav(2017)}]{adepetu2017relative}
\bibinfo{author}{Adepetu, A.}, \bibinfo{author}{Keshav, S.},
  \bibinfo{year}{2017}.
\newblock \bibinfo{title}{The relative importance of price and driving range on
  electric vehicle adoption: Los angeles case study}.
\newblock \bibinfo{journal}{Transportation} \bibinfo{volume}{44},
  \bibinfo{pages}{353--373}.
\bibitem[{Al-Alawi and Bradley(2014)}]{al2014analysis}
\bibinfo{author}{Al-Alawi, B.M.}, \bibinfo{author}{Bradley, T.H.},
  \bibinfo{year}{2014}.
\newblock \bibinfo{title}{Analysis of corporate average fuel economy regulation
  compliance scenarios inclusive of plug in hybrid vehicles}.
\newblock \bibinfo{journal}{Applied energy} \bibinfo{volume}{113},
  \bibinfo{pages}{1323--1337}.
\bibitem[{Ansuj et~al.(1996)Ansuj, Camargo, Radharamanan and
  Petry}]{ansuj1996sales}
\bibinfo{author}{Ansuj, A.P.}, \bibinfo{author}{Camargo, M.},
  \bibinfo{author}{Radharamanan, R.}, \bibinfo{author}{Petry, D.},
  \bibinfo{year}{1996}.
\newblock \bibinfo{title}{Sales forecasting using time series and neural
  networks}.
\newblock \bibinfo{journal}{Computers \& Industrial Engineering}
  \bibinfo{volume}{31}, \bibinfo{pages}{421--424}.
\bibitem[{Bass(1969)}]{bass1969new}
\bibinfo{author}{Bass, F.M.}, \bibinfo{year}{1969}.
\newblock \bibinfo{title}{A new product growth for model consumer durables}.
\newblock \bibinfo{journal}{Management science} \bibinfo{volume}{15},
  \bibinfo{pages}{215--227}.
\bibitem[{Bi(2018)}]{bicheng2018}
\bibinfo{author}{Bi, C.}, \bibinfo{year}{2018}.
\newblock \bibinfo{title}{Analysis of the role of statistical analysis in
  marketing}.
\newblock \bibinfo{journal}{Business Management} \bibinfo{volume}{18},
  \bibinfo{pages}{90--91}.
\bibitem[{Bin et~al.(2013)Bin, Luning and Jianxin}]{renbin@2013diffusionmodel}
\bibinfo{author}{Bin, R.}, \bibinfo{author}{Luning, S.},
  \bibinfo{author}{Jianxin, Y.}, \bibinfo{year}{2013}.
\newblock \bibinfo{title}{A generalized bass model for electric vehicles in
  china based on diffusion of innovation theory}.
\newblock \bibinfo{journal}{Soft Science} \bibinfo{volume}{27},
  \bibinfo{pages}{17--22}.
\bibitem[{Dorigo et~al.(2006)Dorigo, Birattari and Stutzle}]{dorigo2006ant}
\bibinfo{author}{Dorigo, M.}, \bibinfo{author}{Birattari, M.},
  \bibinfo{author}{Stutzle, T.}, \bibinfo{year}{2006}.
\newblock \bibinfo{title}{Ant colony optimization}.
\newblock \bibinfo{journal}{IEEE computational intelligence magazine}
  \bibinfo{volume}{1}, \bibinfo{pages}{28--39}.
\bibitem[{Ellram(1993)}]{ellram1993total}
\bibinfo{author}{Ellram, L.}, \bibinfo{year}{1993}.
\newblock \bibinfo{title}{Total cost of ownership: elements and
  implementation}.
\newblock \bibinfo{journal}{International journal of purchasing and materials
  management} \bibinfo{volume}{29}, \bibinfo{pages}{2--11}.
\bibitem[{Gates(2021)}]{gates2021avoid}
\bibinfo{author}{Gates, B.}, \bibinfo{year}{2021}.
\newblock \bibinfo{title}{How to avoid a climate disaster: the solutions we
  have and the breakthroughs we need}.
\newblock \bibinfo{publisher}{Knopf}.
\bibitem[{Javid and Nejat(2017)}]{javid2017comprehensive}
\bibinfo{author}{Javid, R.J.}, \bibinfo{author}{Nejat, A.},
  \bibinfo{year}{2017}.
\newblock \bibinfo{title}{A comprehensive model of regional electric vehicle
  adoption and penetration}.
\newblock \bibinfo{journal}{Transport Policy} \bibinfo{volume}{54},
  \bibinfo{pages}{30--42}.
\bibitem[{Juster(1966)}]{juster1966consumer}
\bibinfo{author}{Juster, F.T.}, \bibinfo{year}{1966}.
\newblock \bibinfo{title}{Consumer buying intentions and purchase probability:
  An experiment in survey design}.
\newblock \bibinfo{journal}{Journal of the American Statistical Association}
  \bibinfo{volume}{61}, \bibinfo{pages}{658--696}.
\bibitem[{Katoch et~al.(2021)Katoch, Chauhan and Kumar}]{katoch2021review}
\bibinfo{author}{Katoch, S.}, \bibinfo{author}{Chauhan, S.S.},
  \bibinfo{author}{Kumar, V.}, \bibinfo{year}{2021}.
\newblock \bibinfo{title}{A review on genetic algorithm: past, present, and
  future}.
\newblock \bibinfo{journal}{Multimedia Tools and Applications}
  \bibinfo{volume}{80}, \bibinfo{pages}{8091--8126}.
\bibitem[{Kong et~al.(2020)Kong, Xia, Xue and Zhao}]{kong2020effects}
\bibinfo{author}{Kong, D.}, \bibinfo{author}{Xia, Q.}, \bibinfo{author}{Xue,
  Y.}, \bibinfo{author}{Zhao, X.}, \bibinfo{year}{2020}.
\newblock \bibinfo{title}{Effects of multi policies on electric vehicle
  diffusion under subsidy policy abolishment in china: A multi-actor
  perspective}.
\newblock \bibinfo{journal}{Applied energy} \bibinfo{volume}{266},
  \bibinfo{pages}{114887}.
\bibitem[{Kumar et~al.(2022)Kumar, Guha and Chakraborty}]{kumar2022comparative}
\bibinfo{author}{Kumar, R.R.}, \bibinfo{author}{Guha, P.},
  \bibinfo{author}{Chakraborty, A.}, \bibinfo{year}{2022}.
\newblock \bibinfo{title}{Comparative assessment and selection of electric
  vehicle diffusion models: A global outlook}.
\newblock \bibinfo{journal}{Energy} \bibinfo{volume}{238},
  \bibinfo{pages}{121932}.
\bibitem[{Lee et~al.(2017)Lee, Kim and KIM}]{lee2017efficiency}
\bibinfo{author}{Lee, M.h.}, \bibinfo{author}{Kim, S.w.}, \bibinfo{author}{KIM,
  K.H.}, \bibinfo{year}{2017}.
\newblock \bibinfo{title}{The efficiency characteristics of electric vehicle
  (ev) according to the diverse driving modes and test conditions}.
\newblock \bibinfo{journal}{Transactions of the Korean hydrogen and new energy
  society} \bibinfo{volume}{28}, \bibinfo{pages}{56--62}.
\bibitem[{Panwar et~al.(2018)Panwar, Singh and Devadas}]{panwar2018analysis}
\bibinfo{author}{Panwar, M.}, \bibinfo{author}{Singh, D.K.},
  \bibinfo{author}{Devadas, V.}, \bibinfo{year}{2018}.
\newblock \bibinfo{title}{Analysis of environmental co-benefits of
  transportation sub-system of delhi}.
\newblock \bibinfo{journal}{Alexandria engineering journal}
  \bibinfo{volume}{57}, \bibinfo{pages}{2649--2658}.
\bibitem[{Peterson(1989)}]{peterson1989sales}
\bibinfo{author}{Peterson, R.T.}, \bibinfo{year}{1989}.
\newblock \bibinfo{title}{Sales force composite forecasting-an exploratory
  analysis}.
\newblock \bibinfo{journal}{The Journal of Business Forecasting}
  \bibinfo{volume}{8}, \bibinfo{pages}{23}.
\bibitem[{Rogers and Singhal(2003)}]{rogers2003empowerment}
\bibinfo{author}{Rogers, E.M.}, \bibinfo{author}{Singhal, A.},
  \bibinfo{year}{2003}.
\newblock \bibinfo{title}{Empowerment and communication: Lessons learned from
  organizing for social change}.
\newblock \bibinfo{journal}{Annals of the International Communication
  Association} \bibinfo{volume}{27}, \bibinfo{pages}{67--85}.
\bibitem[{Skippon et~al.(2016)Skippon, Kinnear, Lloyd and
  Stannard}]{skippon2016experience}
\bibinfo{author}{Skippon, S.M.}, \bibinfo{author}{Kinnear, N.},
  \bibinfo{author}{Lloyd, L.}, \bibinfo{author}{Stannard, J.},
  \bibinfo{year}{2016}.
\newblock \bibinfo{title}{How experience of use influences mass-market
  drivers’ willingness to consider a battery electric vehicle: a randomised
  controlled trial}.
\newblock \bibinfo{journal}{Transportation Research Part A: Policy and
  Practice} \bibinfo{volume}{92}, \bibinfo{pages}{26--42}.
\bibitem[{Van~Laarhoven and Aarts(1987)}]{van1987simulated}
\bibinfo{author}{Van~Laarhoven, P.J.}, \bibinfo{author}{Aarts, E.H.},
  \bibinfo{year}{1987}.
\newblock \bibinfo{title}{Simulated annealing}, in:
  \bibinfo{booktitle}{Simulated annealing: Theory and applications}.
  \bibinfo{publisher}{Springer}, pp. \bibinfo{pages}{7--15}.
\bibitem[{Wang et~al.(2018)Wang, Zhao, Liu and Hao}]{wang2018impacts}
\bibinfo{author}{Wang, S.}, \bibinfo{author}{Zhao, F.}, \bibinfo{author}{Liu,
  Z.}, \bibinfo{author}{Hao, H.}, \bibinfo{year}{2018}.
\newblock \bibinfo{title}{Impacts of a super credit policy on electric vehicle
  penetration and compliance with china's corporate average fuel consumption
  regulation}.
\newblock \bibinfo{journal}{Energy} \bibinfo{volume}{155},
  \bibinfo{pages}{746--762}.
\bibitem[{Zhu and Du(2018)}]{zhu2018forecasting}
\bibinfo{author}{Zhu, Z.}, \bibinfo{author}{Du, H.}, \bibinfo{year}{2018}.
\newblock \bibinfo{title}{Forecasting the number of electric vehicles: a case
  of beijing}, in: \bibinfo{booktitle}{IOP Conference Series: Earth and
  Environmental Science}, \bibinfo{organization}{IOP Publishing}. p.
  \bibinfo{pages}{042037}.
\bibitem[{Zou et~al.(2020)Zou, Khaloei and MacKenzie}]{zou2020effects}
\bibinfo{author}{Zou, T.}, \bibinfo{author}{Khaloei, M.},
  \bibinfo{author}{MacKenzie, D.}, \bibinfo{year}{2020}.
\newblock \bibinfo{title}{Effects of charging infrastructure characteristics on
  electric vehicle preferences of new and used car buyers in the united
  states}.
\newblock \bibinfo{journal}{Transportation Research Record}
  \bibinfo{volume}{2674}, \bibinfo{pages}{165--175}.

\end{thebibliography}


\end{document}